\documentclass[%
 aip,
 chaos,
 amsmath,amssymb,
 reprint,%
]{revtex4-1}

\usepackage{graphicx}
\usepackage{dcolumn}
\usepackage{bm}

\usepackage[utf8]{inputenc}
\usepackage[T1]{fontenc}
\usepackage{mathptmx}
\usepackage{etoolbox}
\usepackage{xcolor}
\usepackage{hyperref}

\makeatletter
\def\@email#1#2{%
 \endgroup
 \patchcmd{\titleblock@produce}
  {\frontmatter@RRAPformat}
  {\frontmatter@RRAPformat{\produce@RRAP{*#1\href{mailto:#2}{#2}}}\frontmatter@RRAPformat}
  {}{}
}%
\makeatother
\begin{document}

\preprint{AIP/123-QED}

\title{Memory-induced absolute negative mobility}
\author{M. Wi\'{s}niewski}
\author{J. Spiechowicz}%
 \email{jakub.spiechowicz@us.edu.pl}
\affiliation{ 
Institute of Physics, University of Silesia, 41-500 Chorz{\'o}w, Poland
}%


\begin{abstract}
    Non-Markovian systems form a broad area of physics that remains greatly unexplored despite years of intensive investigations. The spotlight is on memory as a source of effects that are absent in their Markovian counterparts. In this work we dive into this problem and analyze a driven Brownian particle moving in a spatially periodic potential and exposed to correlated thermal noise. We show that the absolute negative mobility effect, in which the net movement of the particle is in direction opposite to the average force acting on it, may be induced by the memory of the setup. To explain the origin of this phenomenon we resort to the recently developed effective mass approach to dynamics of non-Markovian systems.
\end{abstract}

\maketitle

\begin{quotation}
Memory is one of the most far-reaching phenomena in nature. It resides at an interdisciplinary crossroad of exact, natural and applied sciences that is equally important from the fundamental as well as practical point of view. On the one hand it constitutes a bridge between  determinism and complete randomness whereas on the other it allows to store digital information as e.g.~magnetic bits. Here we consider its another face and analyze a fundamental problem of the role of memory occurring in a dynamical system. We demonstrate a case in which the memoryless setup is distinctly different than the system with extremely short memory as the latter is the \textit{fons et origo} of a new effect that is absent in the memoryless case.
\end{quotation}

\section{Introduction}
The ultimate goal of theoretical physics is to discover the laws of physics, that is, to establish relationships between the physical quantities. In doing so it very often uses the method of simplification to capture only the essence of the considered system. The consequence of such an approach to e.g.~coupling with the environment results in a number of degrees of freedom that are not directly tracked but emanate as a stochastic perturbation of the system of interest. When the time scale of their dynamics is much shorter than that of the analyzed setup, the Markovian approximation is often employed in which the evolution of the system depends only on its current state but not on the past ones, i.e., the memory is completely neglected.

This assumption, however, is never strictly true.
The time scale of the environment is always finite and therefore it is not entirely negligible.
Moreover, even at the fundamental level of quantum mechanics, the Markovian approximation of a particle coupled to thermal vacuum results in its infinite energy \cite{Spiechowicz2021}.
This means that real systems are never Markovian, or, in other words, they always exhibit memory.
This fact has been reflected in the direction of recent research in fields like active matter \cite{kanazawa2020, banerjee2022, militaru2021, tucci2022}, spin glasses \cite{jesi-baity2023}, protein-folding kinetics \cite{netz}, random walk theory \cite{levenier2022,alessandro2021,guerin2016,barbier2022} and quantum stochastic processes \cite{milz2020, milz2021}, to name only a few.

The non-Markovian systems can differ from their Markovian counterparts not only quantitatively but also qualitatively, which means that they can exhibit phenomena that are absent in their memoryless analogues.
Such effects are then called \textit{memory-induced}.
Their examples include super- and subdiffusion \cite{Kumar2010, goychuk2012, Harbola2014, Goychuk2019}, memory-induced Magnus effect \cite{Cao2023}, memory-induced chaos \cite{Landaw2017} and many more \cite{Glatt2006, Fort2010, Narinder2018, Kappler2018, Mankin2018, Hoffmann2019}.
In this work we analyze the dynamics of a paradigmatic model of statistical physics exhibiting memory, namely a Brownian particle dwelling in a periodic potential and exposed to correlated thermal noise, and report a new phenomenon of the memory-induced absolute negative mobility.

A particle is said to have negative mobility when its net movement is directed opposite to the average force acting on it.
This counterintuitive phenomenon emerges in the world of nonlinear, nonequilibrium physics and its existence has been confirmed both theoretically \cite{McCombie1997, Reimann1999, Cleuren2001, Eichhorn2002prl, Eichhorn2002pre, Cleuren2002, Cleuren2003, Haljas2004, Machura2007, Speer2007pre, Speer2007epl, Kostur2008, Kostur2009, Hanggi2010, Januszewski2011, Du2011, Du2012, Spiechowicz2013jstatmech, Spiechowicz2014pre, Ghosh2014, Malgaretti2014, Dandogbessi2015, Sarracino2016, Slapik2018, Cecconi2018, Ai2018, Cividini2018, Slapik2019prl, Slapik2019prappl, Spiechowicz2019njp, Wu2019, Zhu2019, Luo2020chaos, Luo2020pre, Wisniewski2022, Wu2022, Luo2022, Archana2022, Wisniewski2023} and experimentally \cite{Ros2005, Regtmeier2007epj, Regtmeier2007jss, Nagel2008, Eichhorn2010, Luo2016, Sonker2019}.
Its origin is usually rooted in the complex nonlinear dynamics of the non-equilibrium system, and it can be observed both in zero-temperature deterministic and noisy stochastic setups.
When this phenomenon arises in the regime of linear response, i.e.~for a small biasing force, it is called \textit{absolute} negative mobility.
In this work we present a case in which the absolute negative mobility effect occurs in the system only when (short) memory is taken into account and ceases to exist when the memory vanishes.

The dynamics of our paradigmatic model is described by the Generalized Langevin Equation (GLE), in which memory is characterized by a damping kernel that specifies the correlation of thermal fluctuations.
A common method of the analysis of this equation is the Markovian embedding method, in which the non-Markovian problem is transformed into a Markovian but multidimensional one \cite{Straub1986, Siegle2010}.
Another scheme allows to model the system of interest with a Markovian equation and captures the memory effects as a change in the particle mass, thus it is called the effective mass approach \cite{Wisniewski2024}.
In our work we apply the first method to solve the analyzed equation numerically and make use of the handy interpretation of the latter one to explain the origin of the reported phenomenon.

This article is organized as follows: in the next section we present a model of a Brownian particle in a periodic potential exposed to correlated thermal noise and subjected to an external force. In section \ref{sec:results} we show that for certain model parameters the absolute negative mobility effect arises in the presence of memory but does not occur in the memoryless limit, and explain the origin of this phenomenon with the effective mass approach.
Finally, in section \ref{sec:conclusions} we summarize the work.

\section{Model}

Let us consider a Brownian particle surrounded by a viscoelastic environment and dwelling in a potential $U(x)$ under action of an external force $F_{\mathrm{ext}}(t)$.
Its dynamics can be described with the Generalized Langevin Equation (GLE)
\begin{equation} \label{eq:GLE}
    M\dot{v}(t) + \Gamma \int_{0}^{t} K(t-s)v(s)\mathrm{d}s = -\frac{\mathrm{d}U(x)}{\mathrm{d}x} + F_\mathrm{ext}(t) + \eta(t),
\end{equation}
where $M$ is the particle's mass, $\Gamma$ represents the friction coefficient, $K(t)$ is the memory kernel characterizing relaxation of the system, $F_\mathrm{ext}(t)$ stands for the external force acting on the particle, and $\eta(t)$ represents thermal fluctuations.
The autocorrelation of the latter is related to the kernel $K(t)$ via the fluctuation-dissipation theorem \cite{Kubo1966}
\begin{equation} \label{eq:fd}
    \langle \eta(t)\eta(s) \rangle = \Gamma k_B T K(t-s),
\end{equation}
where $k_B$ is the Boltzmann constant and $T$ stands for the temperature.
Thus, $K(t)$ describes both the temporal correlations of equilibrium fluctuations and the memory experienced by the particle.
In Maxwell's model of viscoelasticity, the damping kernel reads \cite{Maxwell1867, goychuk2012}
\begin{equation} \label{eq:K}
    K(t) = \frac{1}{\tau_c}e^{-t/\tau_c},
\end{equation}
where $\tau_c$ is characteristic time scale describing the correlation time and the memory of the system. The system described by Eq. (\ref{eq:GLE}) has a multitude experimental realizations. Nowadays the most important ones in physics seem to be Josephson junctions \cite{kautz}, SQUIDs \cite{spiechowicz2015njp,spiechowicz2015chaos} and cold atoms dwelling in optical lattices \cite{lutz,barkai}, to mention only a few.

In this work we want to study the directed transport in a nonequlibrium system. For this purpose
we apply to the particle an external force $F_\mathrm{ext}(t) = A\cos(\Omega t) + F$ composed of two parts: (i) an unbiased time-periodic driving $A\cos(\Omega t)$ and (ii) a constant force $F$. The first component drives the system out of thermal equilibrium into a nonequilibrium state but does not induce the directed transport. The latter is caused by the second, static part which breaks the symmetry of the system. Last but not least, we put the particle into a non-confining, spatially periodic and symmetric potential landscape $U(x) = U_0\sin(2\pi x/L)$.
With such a choice of the potential, we define the length and time units as $x_0 = L$ (the spatial period of $U(x)$) and $\tau_0 = \Gamma L^2/V_0$ (the time characterizing relaxation of an overdamped particle in the potential), and operate with the dimensionless variables
\begin{equation}
    \hat{t} = \frac{t}{\tau_0}, \quad \hat{x} = \frac{x}{x_0}, \quad \hat{v} = \frac{v}{v_0},
\end{equation}
where $v_0 = x_0/\tau_0$. Eq.~(\ref{eq:GLE}) can be then transformed into a dimensionless form, which reads
\begin{equation} \label{eq:gle}
    m\dot{\hat{v}}(\hat{t}) + \int_{0}^{\hat{t}} \hat{K}(\hat{t}-s)\hat{v}(s)\mathrm{d}s = -\frac{\mathrm{d}\hat{U}(\hat{x})}{\mathrm{d}\hat{x}} + f_\mathrm{ext}(\hat{t}) + \hat{\eta}(\hat{t}),
\end{equation}
where $m=M/(\Gamma\tau_0)$ and 
\begin{equation}
	\hat{U}(\hat{x}) = \sin(2\pi\hat{x}).
\end{equation}
The dimensionless memory kernel reads
\begin{equation} \label{eq:k}
    \hat{K}(\hat{t}) = \tau_0K(t) = \frac{1}{\tau}e^{-\hat{t}/\tau},
\end{equation}
where $\tau = \tau_c/\tau_0$.
Moreover
\begin{equation}
    f_\mathrm{ext}(\hat{t}) = a\cos(\omega \hat{t}) + f,
\end{equation}
with $a = AL/V_0$, $\omega = \Omega\tau_0$ and $f = FL/V_0$.
The thermal fluctuations $\hat{\eta}(\hat{t})$ obey the relation 
\begin{equation}
    \langle \hat{\eta}(\hat{t}) \hat{\eta}(\hat{s}) \rangle = D\hat{K}(\hat{t}-\hat{s}),
\end{equation}
where $D = k_B T/U_0$, i.e. a ratio of thermal fluctuation energy and half of a periodic potential barrier. 
From now on for clarity we omit the hat over the variables and operate with the dimensionless quantities.

For the rest of the article we consider the following parameter regime
\begin{equation}
    m=0.907,\quad a=12.2,\quad \omega=5.775,\quad D=10^{-4}.
\end{equation}

Since we are interested in the particle mobility, we first define the average velocity in the long time regime as
\begin{equation}
    \langle v \rangle = \lim\limits_{t\to\infty} \frac{1}{t} \int_{0}^{t} \langle \dot{x}(s) \rangle \mathrm{d}s,
\end{equation}
where the brackets $\langle \cdot \rangle$ indicate averaging over the initial conditions and realizations of the thermal noise. The former is required in the deterministic regime where the dynamics may be non-ergodic and consequently the results are affected by the specific choice of initial conditions \cite{spiechowicz2016scirep}.
The mobility $\mu(f)$ characterizes the particle response $\langle v \rangle$ to the static bias $f$ and is defined via the relation \cite{Kostur2008}
\begin{equation}
    \langle v \rangle = \mu(f) f.
\end{equation}
The typical situation occurs when the mobility and its derivative is positive, i.e. $\mu(f) > 0$ and $\mathrm{d}\mu/\mathrm{d}f > 0$, so that the particle moves in the direction of the biasing force and its velocity increases with larger $f$. 
In complex nonlinear systems it happens that the derivative $\mathrm{d}\mu/\mathrm{d}f$ is negative for some interval of $f$, which means that the average velocity decreases when the constant force grows. 
Such phenomenon is known as the \textit{negative differential mobility} \cite{Kostur2008}.
However, there are cases when the particle moves in the direction opposite to the constant force. In such a situation the system exhibits the \textit{negative mobility} $\mu(f) < 0$.
At first glance, this phenomenon seems to contradict the superposition and Le Chatelier–Braun’s principles, but the first constraint does not hold in nonlinear systems, whereas the latter can be violated in systems out of equilibrium \cite{Machura2007, Speer2007pre}.
This means that the negative mobility effect is characteristic to nonlinear, nonequilibrium systems.
In its most astonishing variant, the negative mobility effect occurs in the linear response regime for which it no longer depends on the external static bias
\begin{equation}
    \mu_0 = \lim\limits_{f\to0} \mu(f).
\end{equation}
The case $\mu_0 < 0$ is known as the \textit{absolute negative mobility} (ANM) \cite{Machura2007}.

Eq.~(\ref{eq:gle}) is a stochastic integro-differential nonlinear equation which cannot be solved analytically.
In order to investigate the particle dynamics numerically, we used the Markovian embedding technique, i.e.~extended the phase space of the system so that it obeys a (time-local) Markovian equation \cite{Straub1986}. In doing so we transformed Eq.~(\ref{eq:gle}) into an equivalent set of first-order differential equations
\begin{subequations} \label{eq:gle_set}
\begin{align}
    \dot{x}(t) &= v(t), \\
    m\dot{v}(t) &= -U'(x) + f_{\mathrm{ext}}(t) + z(t), \\
    \tau\dot{z}(t) &= -z(t) - v(t) + \xi(t),
\end{align}
\end{subequations}
where $z(t)$ is the auxiliary embedding variable and $\xi(t)$ is white thermal noise obeying the relation
\begin{equation}
	\langle \xi(t) \xi(s) \rangle = 2D\delta(t-s).
\end{equation}
Then, to numerically solve the set of equations (\ref{eq:gle_set}) we employed a weak second-order predictor-corrector algorithm \cite{Platen2010}. To speed up the necessary calculations, we implemented the algorithm in the CUDA environment on modern graphics processing units, which allowed us to accelerate the simulations by several orders of magnitude \cite{Spiechowicz2015}.

\section{Results} \label{sec:results}

\begin{figure}[t]
    \centering
    \includegraphics{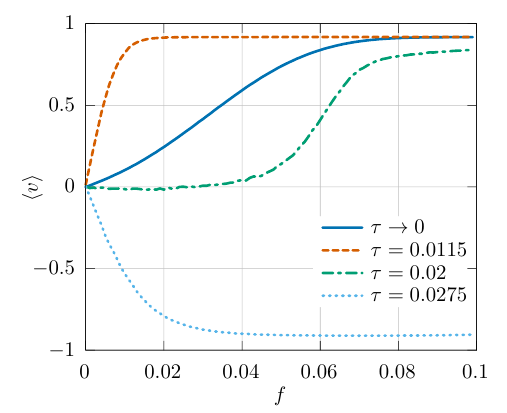}
    \caption{The average velocity $\langle v \rangle$ of the Brownian particle determined from the GLE (\ref{eq:gle}) as a function of the static force $f$ for different values of the memory time $\tau$.}
    \label{fig:v(f)}
\end{figure}

We start our discussion of the results with Fig. \ref{fig:v(f)}, in which we show the average velocity $\langle v \rangle$ versus the static force $f$ for different values of the memory time $\tau$. 
The mobility in the linear response regime $\mu_0$ can be interpreted as the slope of the $\langle v \rangle(f)$ curve for vanishing bias $f \to 0$. We observe that in the Markovian limit $\tau \to 0$ of white thermal noise the mobility is positive $\mu_0 > 0$. As $\tau$ grows, the slope initially becomes steeper, however, for the correlation time \mbox{$\tau = 0.0275$} corresponding to non-Markovian dynamics it is clearly negative $\mu_0 < 0$ leading to the ANM phenomenon. It means that in the considered parameter regime this effect emerges only for the system with the non-zero memory time.

\begin{figure}[t]
    \centering
    \includegraphics{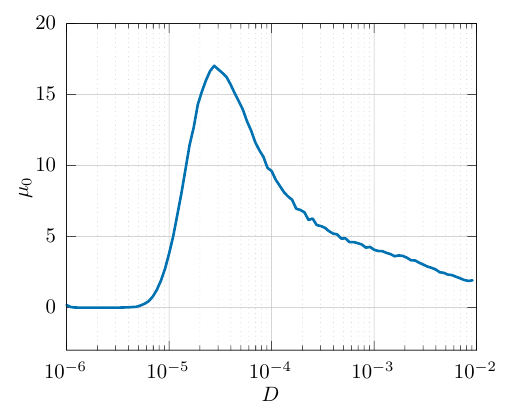}
    \caption{The mobility $\mu_0$ of the Brownian particle in the linear response regime versus temperature $D$ in the Markovian (memoryless) limit $\tau \to 0$.}
    \label{fig:mu0(D)}
\end{figure}

In Fig. \ref{fig:mu0(D)} we depict the mobility $\mu_0$ of the Brownian particle in the linear response regime as a function of temperature $D$ of the system for the Markovian limit of white thermal noise $\tau \to 0$. 
It turns out that the mobility is non-negative $\mu_0 \geq 0$ for all temperatures, including the deterministic setting $D = 0$. It means that in the parameter regime studied in Fig. \ref{fig:v(f)}, the ANM effect is not rooted in the deterministic dynamics of the system, nor can it be caused by white thermal fluctuations. This observation implies that this paradoxical response of the system is induced purely by correlations of thermal equilibrium fluctuations or, equivalently, memory effects in the non-Markovian dynamics.

To the best of our knowledge, the fact that the ANM can be generated solely by (very small!) memory has never been reported before. In [\onlinecite{Kostur2009}] and [\onlinecite{Cecconi2018}] the authors detect the emergence of negative mobility as a result of correlations of thermal fluctuations but in the nonlinear response regime, which is generally vastly more populated in the parameter space of the model than the absolute one. On the other hand, in [\onlinecite{Cleuren2002}] the ANM is induced by memory in the phenomenological model of modified random walk. However, an inspection of Eq.~(4) therein reveals that the directed transport vanishes identically in the memoryless system regardless of the magnitude of applied force, and therefore it is not a correct model. In contrast, in our fully microscopic approach, the presence of a tailored amount of memory reverses the mobility $\mu_0$ in the linear regime from positive to negative.

The origin of the memory-induced ANM in the system described by GLE (\ref{eq:gle}) can be understood if we apply the effective mass approach \cite{Wisniewski2024} to our model.
The general idea behind it is that if the correlation time $\tau$ is short
the GLE can be approximated with a much simpler equation
\begin{equation} \label{eq:gle_approx}
    m^*\dot{v}(t) + v(t) = -\frac{dU(x)}{dx} + f_{\mathrm{ext}}(t) + \xi(t),
\end{equation}
where $m^* = m - \Delta m$ is the effective mass of the particle.
The mass correction is defined as
\begin{equation} \label{eq:dm}
    \Delta m = \int_{0}^{\infty} tK(t)\mathrm{d}t
\end{equation}
and is a function of the correlation time $\tau$.
For the kernel specified by Eq.~\ref{eq:k}, the mass correction reads $\Delta m = \tau$.
Although Eq.~(\ref{eq:gle_approx}) does not involve memory directly and the corresponding pair $\{x, v\}$ forms a Markovian process, the memory effects are reflected in the mass correction $\Delta m$.
The appealing interpretation stating that the short memory is equivalent to a mass shift in the memoryless system allows us to explain the origin of the memory-induced ANM effect.

\begin{figure}[t]
    \centering
    \includegraphics{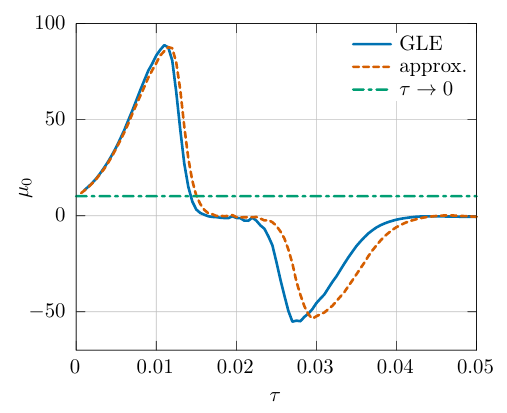}
	\caption{The mobility $\mu_0$ of the Brownian particle in the linear response regime versus the memory time $\tau$ for the original (Eq. (\ref{eq:gle}), solid blue line) and approximate (Eq. (\ref{eq:gle_approx}), dashed red line) system. Dashed-dotted green line illustrates the naive Markovian approximation of white thermal noise without renormalization of the Brownian particle mass. Other parameters are the same as in Fig. \ref{fig:v(f)}.}
    \label{fig:mu0(tau)}
\end{figure}

In Fig. \ref{fig:mu0(tau)}, we present the mobility $\mu_0$ as a function of the memory time $\tau$ for the original (Eq. (\ref{eq:gle})) and approximate (Eq. (\ref{eq:gle_approx})) system. 
For $\tau < 0.01$, the characteristics are barely distinguishable. When $\tau$ increases, the approximate curve starts to ``lag'' behind the original one, but the qualitative behavior remains the same. Most importantly, for both the original and approximate models, mobility $\mu_0$ rises from the positive value, reaches the maximum, passes through zero, and then takes negative values, indicating the ANM effect. 

We note that in the presented case the memory time is \emph{two orders} of magnitude smaller than other characteristic time scales in the system. Indeed, the dimensionless Langevin time of the particle reads $\tau_L = (M/\Gamma)/\tau_0 = m = 0.907$ and the period of the driving force equals $\mathsf{T} = 2\pi/\omega = 1.088$. At first glance, one could think that then the memory effects could be negligible and therefore naively apply the Markovian approximation of white thermal noise without renormalization of the Brownian particle mass (i.e. $m^*=m$ in Eq.~(\ref{eq:gle_approx})). As the dashed-dotted green line in Fig. \ref{fig:mu0(tau)} demonstrates, even in this extreme scenario, it completely fails to correctly predict the system behavior. In particular, for $\tau \to 0$ the mobility $\mu_0$ of the Brownian particle is only positive in contrast to the actual solution when the memory-induced ANM effect emerges. Moreover, even for the extremely short memory $\tau = 0.01$ for which $\mu_0$ is positive, the difference between the mobility corresponding to the real value and the naive Markovian approximation is almost an order of magnitude.

\begin{figure}[t]
    \centering
    \includegraphics{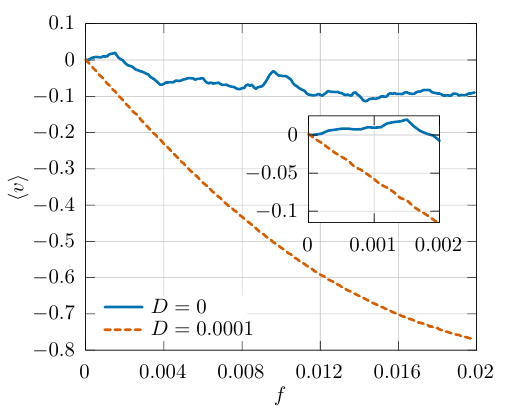}
    \caption{The dimensionless average velocity $\langle v \rangle$ of the Brownian particle as a function of the static force $f$ for $m=0.8775$ (the minimum of the approximate $\mu_0(\tau)$ characteristic, see Fig. \ref{fig:mu0(tau)}) for the approximate system given by Eq. (\ref{eq:gle_approx}). Other parameters are the same as in Fig. \ref{fig:v(f)}.}
    \label{fig:v(f)_m}
\end{figure}

Finally, Fig. \ref{fig:v(f)_m} shows the average velocity $\langle v \rangle$ of the Brownian particle in the approximation (\ref{eq:gle_approx}) versus the static force $f$ for $\tau=0.0295$ which corresponds to the minimum of $\mu_0(\tau)$ in the effective mass approach, see Fig. \ref{fig:mu0(tau)}. 
In the deterministic case $D=0$, the mobility in the limit $f \to 0$ is positive. In contrast, for temperature $D=0.0001$, it is negative. It means that in the approximate Markovian dynamics (\ref{eq:gle_approx}) the absolute negative mobility is caused by white thermal fluctuations \cite{Machura2007} but for the renormalized mass of the particle.

\section{Conclusions} \label{sec:conclusions}

In this work we analyzed non-Markovian dynamics of a Brownian particle dwelling in a periodic potential and driven by both a time-periodic force and a constant bias. The memory of the system is characterized by the correlation time $\tau$ of thermal fluctuations. We revealed a set of the system parameters for which the absolute negative mobility effect, namely the situation when the particle moves in the direction opposite to the applied small static bias, is induced by the memory.

Specifically, the mobility $\mu_0$ of the particle in the linear response regime stays positive for any temperature $D$ of the system in the memoryless limit $\tau \to 0$. When $\tau$ increases, initially so does the mobility, however, after reaching a maximum it decreases and finally takes negative values, indicating the absolute negative mobility effect. It means that this phenomenon is induced by the memory of the system.

To explain the origin of this effect we employed the recently developed effective mass approach to memory in non-Markovian systems. According to this method the presence of short memory in the system is equivalent to renormalization of its mass in the memoryless counterpart of the setup. Indeed, we revealed that the absolute negative mobility effect emerges in the system without memory but with the reduced mass. The phenomenon in this shifted regime is induced by thermal fluctuations.

From a more general perspective this result shows that even the extremely short memory can radically change behavior of the system and lead to unexpected and paradoxical phenomena such as the memory-induced absolute negative mobility effect. This fact must be contrasted with common reasoning that if the memory is extremely small the memoryless (Markovian) approximation can be safely used. We show that even in such a case special caution may be needed. 

\section*{Acknowledgment}
This work was supported by the Grant NCN 2022/45/B/ST3/02619 (J.S.)

\section*{Conflict of interest}
The authors have no conflicts to disclose.

\section*{Data availability statement}
The data that support the findings of this study are available from the corresponding author upon reasonable request.

\section*{References}


\end{document}